\documentclass[conference]{IEEEtran}
\IEEEoverridecommandlockouts
\usepackage{cite}
\usepackage{amsmath,amssymb,amsfonts}
\usepackage{algorithmic}
\usepackage{graphicx}
\usepackage{textcomp}
\usepackage{xcolor}
\usepackage{amsmath,graphicx}
\usepackage{mathtools, amsfonts, url, cite, hyperref, subcaption, booktabs, caption}
\usepackage{subfloat}

\def\BibTeX{{\rm B\kern-.05em{\sc i\kern-.025em b}\kern-.08em
    T\kern-.1667em\lower.7ex\hbox{E}\kern-.125emX}}
\begin{document}

\title{VC-ENHANCE: Speech Restoration with Integrated Noise Suppression and Voice Conversion\\
% {\footnotesize \textsuperscript{*}Note: Sub-titles are not captured for https://ieeexplore.ieee.org  and
}

\author{\IEEEauthorblockN{Kyungguen Byun}
\IEEEauthorblockA{\textit{Qualcomm Technologies Inc.} \\
San Diego, CA \\
kyunggue@qti.qualcomm.com}
\and
\IEEEauthorblockN{Jason Filos}
\IEEEauthorblockA{\textit{Qualcomm Technologies Inc.} \\
San Diego, CA \\
jfilos@qti.qualcomm.com}
\and
\IEEEauthorblockN{Erik Visser}
\IEEEauthorblockA{\textit{Qualcomm Technologies Inc.} \\
San Diego, CA \\
evisser@qti.qualcomm.com}
\and
\IEEEauthorblockN{Sunkuk Moon}
\IEEEauthorblockA{\textit{Qualcomm Technologies Inc.} \\
San Diego, CA \\
sunkukm@qti.qualcomm.com}
}

\maketitle

\begin{abstract}
Noise suppression (NS) algorithms are effective in improving speech quality in many cases. However, aggressive noise suppression can damage the target speech, reducing both speech intelligibility and quality despite removing the noise. This study proposes an explicit speech restoration method using a voice conversion (VC) technique for restoration after noise suppression. We observed that high-quality speech can be restored through a diffusion-based voice conversion stage, conditioned on the target speaker embedding and speech content information extracted from the de-noised speech. This speech restoration can achieve enhancement effects such as bandwidth extension, de-reverberation, and in-painting. Our experimental results demonstrate that this two-stage NS+VC framework outperforms single-stage enhancement models in terms of output speech quality, as measured by objective metrics, while scoring slightly lower in speech intelligibility. To further improve the intelligibility of the combined system, we propose a content encoder adaptation method for robust content extraction in noisy conditions.
\end{abstract}


\begin{figure*}
\includegraphics[width=\textwidth]{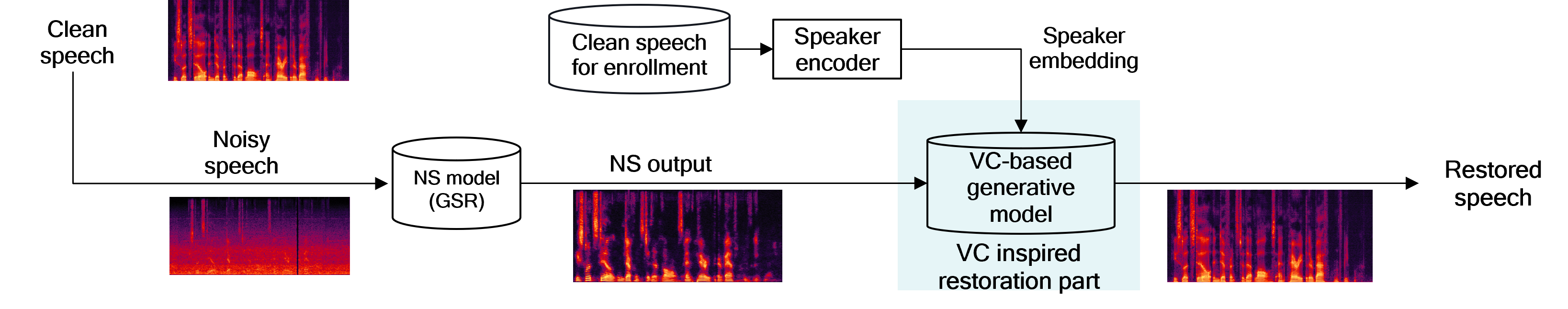}
\centering
\caption{\label{first} Inference process of the proposed system.} %--> Fig.1a}
\end{figure*}

\begin{figure}
\includegraphics[width=\linewidth]{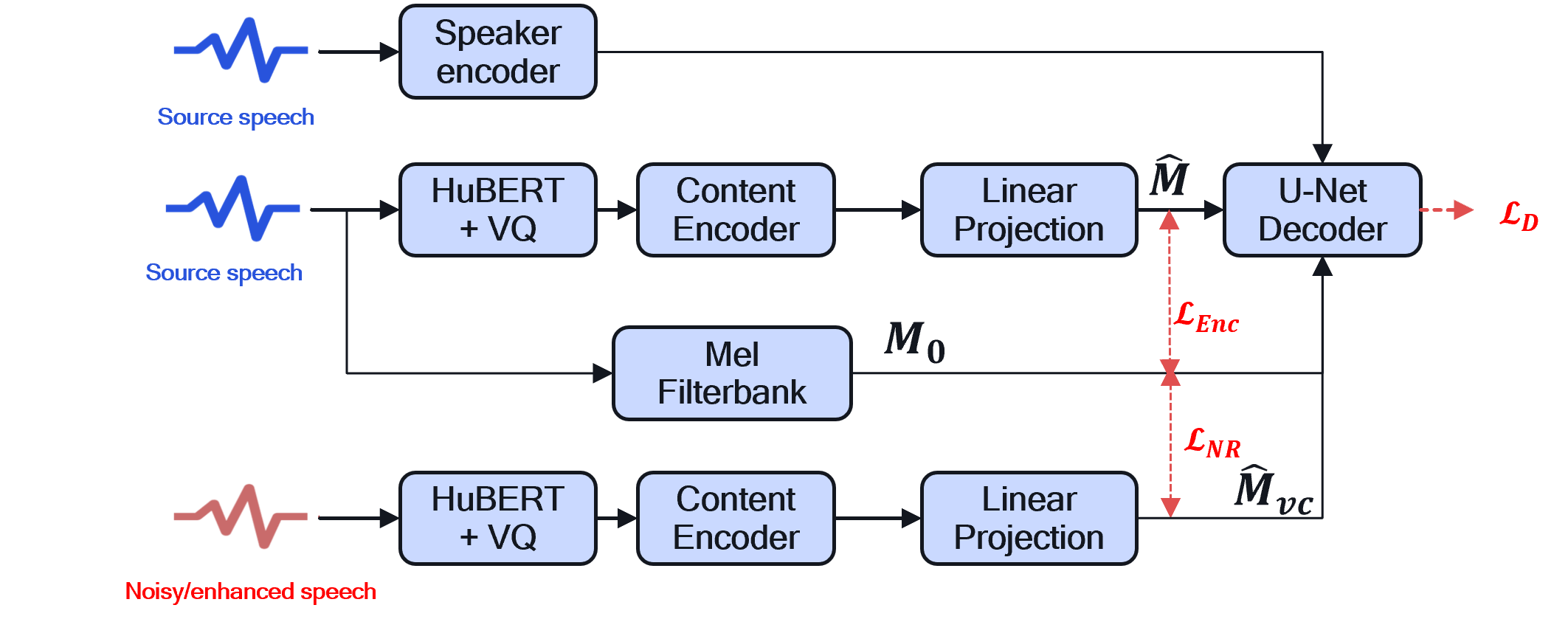}
\centering
\caption{\label{second} Training and content encoder adaptation framework for voice conversion-based speech restoration model.} %--> Fig.1b
\end{figure}

\section{Introduction}
Speech enhancement, also known as noise suppression (NS), is crucial in real-world applications such as telecommunications, voice assistants, and hearing aids. One notable limitation of NS systems is their performance in challenging signal-to-noise ratio (SNR) scenarios, where the enhanced speech often ends up being muffled or choppy \cite{wang2018supervised}. Consequently, NS users may experience reduced speech intelligibility and increased distortion, particularly with challenging acoustic distortions such as in clipped, band-limited, and reverberant speech.

Speech restoration is an alternative approach to improving speech quality. Unlike noise suppression (NS), which aims to attenuate undesired noise, speech restoration endeavors to recover the original speech signal from its degraded form \cite{liu2022voicefixer}\cite{zhang2021restoring}. Consequently, in scenarios where the original data is either completely lost or severely corrupted, such as in cases of clipping or packet loss, speech restoration may offer a better solution. Generative models, such as generative adversarial networks (GANs) \cite{goodfellow2020generative} or diffusion probabilistic models \cite{song2020score}, are typically used for this purpose instead of masking models in NS.

Voice conversion (VC) is a promising method for speech restoration, aiming to transform speech from a source speaker to a target speaker while preserving linguistic content. Recent VC models \cite{autovc}\cite{freevc}\cite{kaneko2020cyclegan}\cite{diffvc} extract speaker-independent content features, such as phonetic posteriorgrams, from the input speech. Generative models then use these features to produce target speech with speaker-specific characteristics, like speaker embeddings. The training process for VC is similar to that of speech restoration, as it involves extracting speaker-specific features from the source speech and using a generative step to restore the original speech state under appropriate guidance. Typically trained on clean data, VC models can generate high-quality outputs. However, they are susceptible to corruption when the input speech is noisy, as they are not usually trained with noisy data.

In this paper, we propose an integrated framework that combines speech enhancement and restoration. Our framework includes an NS module followed by a restoration module. The NS stage is based on \cite{liu2022voicefixer}, while the restoration stage is inspired by the Diff-VC method \cite{diffvc}. In the restoration stage, we condition the generative model with a clean speech speaker embedding from the same speaker as the speech to be de-noised. We assume short, uncorrelated segments of clean speech are obtained beforehand. We modified Diff-VC by replacing the content encoder with the vector-quantized output of the HuBERT encoder for better linguistic feature extraction. Additionally, content encoder adaptation is employed to efficiently integrate NS and VC models.

To our knowledge, the proposed speech enhancement and restoration system offers several novel contributions. \textbf{Superior speech quality}: By combining speech restoration and enhancement, our system achieves better speech quality than single-stage NS methods. \textbf{Robust speaker preservation}: Using voice conversion techniques as a restoration module allows us to restore degraded speech signals while preserving the speaker’s characteristics. \textbf{Integrated system model adaptation}: We propose a model adaptation technique to integrate the two modules by fine-tuning the restoration module on the NS module outputs for robust speech content extraction.

\section{Background}
\subsection{Diffusion models}
Diffusion models have shown extraordinary performance in generative tasks and are utilized in speech related applications \cite{borsos2023audiolm}\cite{serra2022universal}\cite{gradtts}. The diffusion process, a stochastic method frequently used in generative modeling, consists of two primary stages: the forward process and the backward process \cite{song2020score}. In the forward process, noise is incrementally added to the data, transforming it into a noise distribution. The backward process aims to reconstruct the original data from the noisy data by reversing the forward process. During the reverse process, the model is trained to learn a scoring function that represents the gradient of the log probability density \cite{song2020score}. This framework allows for the generation of high-fidelity data by iteratively denoising the noisy data, making it a powerful tool in various applications.

\subsection{HuBERT (Hidden-Unit BERT) model}
HuBERT (Hidden-Unit BERT)\cite{hubert}\footnote{https://github.com/facebookresearch/textlesslib} is a self-supervised speech representation learning model designed to extract informative features without a lexicon during pre-training. The HuBERT model is trained with a BERT-like \cite{devlin2018bert} prediction loss, focusing on masked regions to learn a combined acoustic and language model. Additionally, HuBERT employs an offline clustering step to generate discrete units, allowing it to transform continuous speech inputs into discrete hidden units and learn a robust text-related speech representation.

\section{Method}
An overview of our proposed system is shown in Fig. 1. The noisy speech is first processed by the speaker-agnostic NS model. Then, a VC-inspired restoration model generates restored speech by conditioning the diffusion process with a clean speech target speaker embedding. The clean speech used for extracting the speaker embedding does not need to be content aligned with the noisy speech.

\label{sec:proposed}
\subsection{Noise suppression}
Our noise suppression model is based on the ResU-Net structure used in VoiceFixer \cite{liu2022voicefixer}. VoiceFixer is a comprehensive framework designed for high-fidelity speech restoration, capable of addressing multiple types of distortions such as noise, reverberation, and clipping. The model consists of a two-stage process: the analysis stage and the synthesis stage. In the analysis stage, it predicts enhanced intermediate-level speech features from the degraded input speech, and in the synthesis stage, it generates the waveform using a neural vocoder. There are two key differences between VoiceFixer and our implementation. The first is that we used  HiFi-GAN \cite{hifigan}, which takes a 64-dimensional Mel-spectrogram as input in the synthesis stage. The second difference is that we added the restoration output to the mixture of the system output, instead of using a masked-based multiplicative approach.

\subsection{VC inspired speech restoration}

The VC-based speech restoration module, as shown in Fig. 2, is based on a variation of the Diff-VC \cite{diffvc} model. We use the pre-trained HuBERT encoder and vector quantization (VQ) with a cluster size of 2000 for content embedding, instead of the phoneme-wise spectrogram averaging method in Diff-VC. Then, we employ a transformer-based content encoder and a linear projection layer to map the discrete HuBERT+VQ output to a coarse spectrogram. Next, using a U-Net decoder in the diffusion process, which takes the coarse spectrum and speaker embedding extracted from the speaker encoder, we obtain a restored spectrogram that closely matches the target. Finally, the pre-trained HiFi-GAN \cite{hifigan}\footnote{https://huggingface.co/speechbrain/tts-hifigan-libritts-16kHz} is used to generate the speech waveform from the U-Net decoder output.

The VC model is trained using a weighted sum of two loss functions as shown in Eq. (1). The encoder loss, $L_{enc}$, minimizes the L1-distance between $\bold{M}_{0}$, and $\bold{\hat{M}}$, the clean Mel-spectrogram, and the output of the content encoder after linear projection. The diffusion loss, $L_{d}$, is used to train the U-Net decoder part, as shown in Eq. (3).
\begin{equation}
L_{total} = L_{d} + \alpha L_{enc}
\end{equation}
\begin{equation}
L_{enc}(\bold{M_0}, \hat{\bold{M}}) = \mathbb{D}(\bold{M_0}, \hat{\bold{M}}),  
\end{equation}
\begin{equation}
L_d(\bold{M_{0}}) = \mathbb{E}_{\epsilon_t}[|| s_\theta(M_t, t | \hat{M}, \bold{s}) + \epsilon_{t}||^{2}_{2}].
\end{equation}
Here, $M_{t}$, and $\epsilon_{t}$ represent the Mel-spectrogram and noise at time step $t$, respectively. The score function, $s_\theta$, predicts the scale of the noise $\epsilon_{t}$ conditioned on the speaker embedding $\bold{s}$ and content embedding, $\bold{\hat{M}}$. Classifier-free guidance \cite{guidance} is used during training with a probability of $p=0.1$ for speaker and content embeddings. The parameter $\alpha$ is set to 0.1. 

\subsection{Content encoder adaptation}
In order to ensure accurate speech content extraction by the VC stage, the content encoder in Fig 2 was adapted to help it understand the enhanced speech output from the NS stage. This is necessary as clean speech is typically used during VC model training and enhanced speech may contain distortions and noise residuals unseen during training.  A noise robust loss, $L_{nr}$, is introduced: $L_{nr}(\bold{M_0}, \hat{\bold{M}}_{vc}) = \mathbb{D}(\bold{M_0}, \hat{\bold{M}}_{vc})$, where $\hat{\bold{M}}_{vc}$ is output of the content encoder after linear projection from noisy and enhanced speech. The $L_{nr}$ is merged as total loss function: $L_{total} = L_{d} + \alpha( L_{enc} + L_{nr})$.

\begin{table*}
    \caption{Experimental results: Noisy mixtures, NS output, proposed (VC-E), proposed w/ adaptation (VC-aE)}
    \centering
    \begin{tabular}{ccccccccccccc}
        \toprule
        \textbf{Metric} & \multicolumn{4}{c}{\textbf{NISQA ($\uparrow$)}} & \multicolumn{4}{c}{\textbf{SECS ($\uparrow$)}} & \multicolumn{4}{c}{\textbf{CER [\%] ($\downarrow$)}} \\
        \midrule
        \textbf{Method} & \textbf{Noisy} & \textbf{NS} & \textbf{VC-E} & \textbf{VC-aE} & \textbf{Noisy} & \textbf{NS} & \textbf{VC-E} & \textbf{VC-aE} & \textbf{Noisy} & \textbf{NS} & \textbf{VC-E} & \textbf{VC-aE} \\
        \midrule
        \textbf{Clean} & 4.05 & - & - & - & - & - & - & - & 0.41 & - & - & - \\
        \textbf{25dB} & 3.32 & 3.63 & 3.75 & 3.95 & 0.55  & 0.54 & 0.60 & 0.52 & 0.57 & 1.15 & 11.81 & 6.68\\
        \textbf{20dB}  & 3.08 & 3.62 & 3.77 & 3.93 & 0.53 & 0.53  & 0.59  & 0.52 & 0.87 & 1.21 & 10.06 & 7.28\\     
        \textbf{15dB} & 2.86 & 3.55 & 3.74 & 3.95 & 0.51 & 0.51 & 0.59 & 0.51 & 0.79 & 2.69 & 13.00 & 9.15\\ 
        \textbf{10dB} & 2.57 & 3.49 & 3.74 & 3.93 & 0.47 & 0.48 & 0.57 & 0.50 & 0.89 & 7.68 & 24.98 & 14.88\\
        \textbf{5dB} & 2.17 & 3.43 & 3.74 & 3.98 & 0.44 & 0.45 & 0.54 & 0.50 & 2.45 & 10.51 & 32.06 & 21.69\\       
        \midrule
        \textbf{Avg.} & 3.01 & 3.54 & 3.75 & 3.95 & 0.50 & 0.50 & 0.58 & 0.51 & 1.00 & 4.65 & 18.38 & 11.94\\       
        \bottomrule
    \end{tabular}
    \label{tab:experiment_result}
\end{table*}

\begin{table*}[t]
\caption{CER 12\% example}
\centering
\begin{tabular}{cc}
\toprule
\textbf{Ground Truth} & Throughout the centuries people have explained the rainbow in various ways. \\
\midrule
\textbf{Transcribed} & Throughout the centuries, people have exclaimed to Abar in various ways. (12.73\%) \\
\midrule
\textbf{Ground Truth} & Others have tried to explain the phenomenon physically.  \\
\midrule
\textbf{Transcribed} & Others will try to explain the phenomenon physically. (12.00\%) \\
\bottomrule
\end{tabular}
\end{table*}

\begin{table}[t]
\caption{STOI scores}
\centering
\begin{tabular}{ccccc}
\toprule
\textbf{Metric} & \multicolumn{4}{c}{\textbf{STOI ($\uparrow$)}}   \\
\midrule
\textbf{Method}  & \textbf{Noisy} & \textbf{NS} & \textbf{VC-E} &\textbf{VC-aE} \\
\midrule
\textbf{25dB}   & 0.96 & 0.90 & 0.71 & 0.69  \\
\textbf{20dB}   & 0.94 & 0.89 & 0.71 & 0.69  \\
\textbf{15dB}   & 0.91 & 0.88 & 0.70 & 0.68  \\
\textbf{10dB}   & 0.87 & 0.85 & 0.69 & 0.68  \\
\textbf{5dB}    & 0.81 & 0.83 & 0.68 & 0.66  \\
\midrule
\textbf{Avg.}   & 0.90 & 0.87 & 0.70 & 0.68  \\
\bottomrule
\end{tabular}
\end{table}

\section{Experiments}
\label{sec:experiment}
\subsection{Evaluation metrics \& datasets}
To evaluate the performance of the proposed systems, we employed four objective metrics. The NISQA \cite{nisqa}\footnote{https://github.com/gabrielmittag/NISQA} score predicts the mean opinion score of the input speech. The speaker embedding cosine similarity (SECS) measures the speaker similarity. The character error rate (CER) \cite{gordeeva2021meaning} assesses the accuracy of the speech transcription. Finally, the STOI \cite{taal2010short} metric evaluates the intelligibility of the system. We used a pre-trained ECAPA-TDNN \cite{ecapa} model to extract speaker embeddings. Whisper Large-v2 \cite{whisper}\footnote{https://huggingface.co/openai/whisper-large-v2} was used to obtain the speech transcript and measure the CER.
% \footnote{https://huggingface.co/openai/whisper-large-v2}
% \footnote{https://github.com/gabrielmittag/NISQA}

\begin{figure}[t]
\begin{subfigure}{\columnwidth}
\centering
\includegraphics[width=0.8\textwidth]{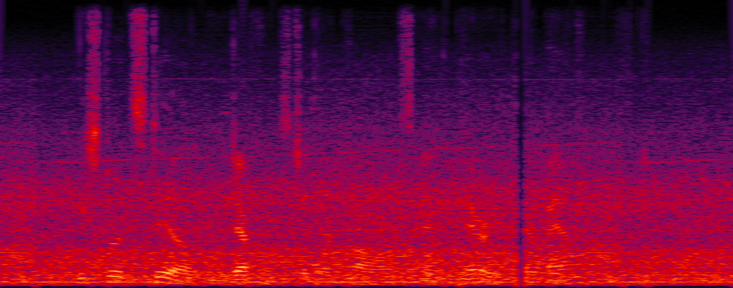}
\caption{\label{first} Noisy input spectrogram.} %--> Fig.1a}
\end{subfigure}
\begin{subfigure}{\columnwidth}
\centering
\includegraphics[width=0.8\textwidth]{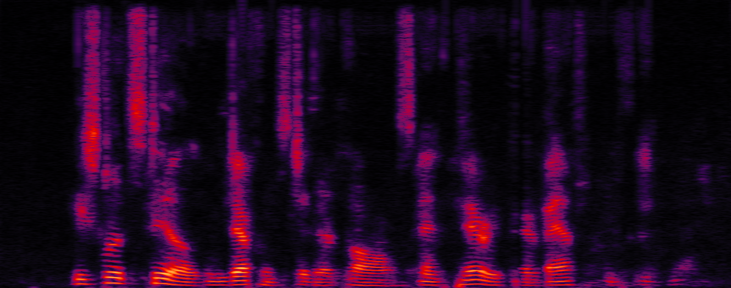}
\caption{\label{second} Noise suppression model output.}
\centering
\end{subfigure}
\begin{subfigure}{\columnwidth}
\centering
\includegraphics[width=0.8\textwidth]{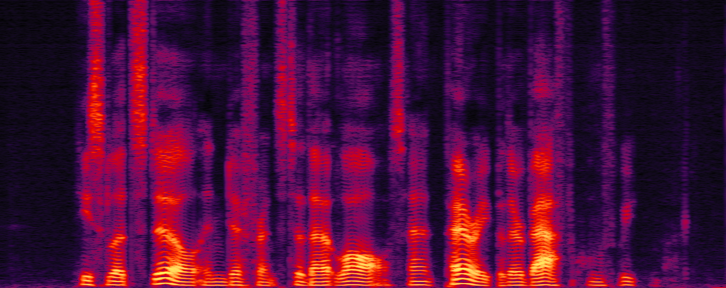}
\centering
\caption{\label{third} Proposed model output.}
\end{subfigure}
\caption{Input mixture including noise and packet-loss and associated outputs.}
\label{overall}
\end{figure}

The voice conversion model is trained in a 2-stage manner. First, we trained on a clean dataset, including train-clean-100h and 360h subsets of LibriTTS \cite{libritts}, for 100 epochs. For the content encoder adaptation method, we prepared a noisy dataset by mixing various types of noise with clean speech using an SNR in $\{10 \textrm{dB}, 20 \textrm{dB}\}$. To evaluate model performance, we prepared clean and noisy samples with SNRs in $\{5 \textrm{dB}, 25 \textrm{dB}\}$ from the VCTK \cite{vctk} dataset, ensuring these samples were different from those used during training.

\subsection{Experimental result}
We consider 4 cases  for comparison (Table 1), \textbf{Noisy} represents results obtained for noisy mixtures without processing, \textbf{NS} represents NS-only outputs, \textbf{VC-E} and \textbf{VC-aE} represent the proposed system, VC-Enhance, without and with the content encoder adaptation, respectively.

Table 1 summarizes the performance obtained for various SNRs. We observed an increase in NISQA scores for all systems (NS, VC-E, VC-aE) compared to the unprocessed case, which shows gradually decreasing NISQA scores with lower SNR. The proposed system, VC-aE, achieves the highest average score (3.95), comparable to that of clean speech (4.05). Notably, the NISQA score is maintained at a similar level across SNR conditions with the proposed approaches, compared to the NS-only case. This suggests a high-quality speech generation capability of the diffusion decoder.

With respect to speaker similarity as measured by SECS, we found that the VC-E systems outperform the NS-only model. The VC restoration module manages to use the clean speech speaker embedding conditioned diffusion model to regenerate target speech characteristics and undo the damage incurred in the NS stage. We observed a slight degradation of SECS in the content embedding adaptation VC-aE case, but it is still better than the NS only system on average. 

Regarding intelligibility, higher CER is observed for all processing stages compared to no processing. The lowest average CER, 4.65\%, is observed with the NS-only stage, while the VC-E system shows 18.38\%. With the content encoder adaptation technique VC-aE, the average CER decreases to 11.94\%, a 35\% relative reduction compared to the VC-aE system. This suggests that the 2-stage VC approaches significantly degrade intelligibility compared to unprocessed/NS-only cases, and that speech content extraction is not robust enough in noisy conditions, even after content encoder adaptation. To investigate the degree of content corruption, Table 2 shows examples of CER around 12\%, similar to the average CER of the proposed method. Most content words are correctly transcribed; however, consonants in several words are replaced, indicating that misclassification of VQ indices occurs more often for consonants in low SNR conditions.

To provide an alternative measure of intelligibility, STOI \cite{taal2010short} scores are shown in Table 3. GCC-PHAT \cite{knapp1976generalized} was used to synchronize the clean and restoration system outputs to compute the STOI scores. The STOI of the proposed VC-aE method is around 0.7, slightly lower than that of the Noisy or NS system, but still considered acceptable as good quality \cite{Dong2020}, \cite{Taal2016}. Given this STOI, the qualitative examples in Table 2, and the CER results in Table 1, it is suggested that CER is too sensitive to measure intelligibility and should be considered more as a measure of how suited the proposed approach is for human listeners or ASR downstream tasks.

In summary, the proposed 2-stage approaches provide significant speech quality improvement compared to NS-only system, while incurring a slight degradation in intelligibility.

\subsection{Ablation study}
We conducted several ablation studies to identify which parts of the model are critical for performance. First, we removed VQ to determine whether the dense HuBERT features help extract robust content embeddings. We found that a lower CER was obtained without VQ; however, this was accompanied by a severe drop in NISQA scores. Without VQ, noise components leaked into the content encoder, resulting in noisy Mel-spectrograms. We also experimented with a VC-based restoration-only system (no NS). This system achieved lower CER in high SNR cases but suffered from high CER in low SNR conditions, yielding worse performance on average. Hence, we concluded that NS pre-processing for the VC model helps make content embedding extraction more robust.

\subsection{Testing in various adverse conditions}
We introduced three types of artifacts to test the restoration process: background noise, low- \& high-pass filtering, and packet-drop. The latter means that entire regions of the waveform are masked for a short period of time (approximately 100 ms). An example Mel-spectrogram is shown in Figure 3. After the NS stage, a less noisy speech is obtained, but it contains many areas of muffled and choppy speech. In the proposed VC-aE output, we see that the restoration can handle multiple artifacts simultaneously. A better harmonic structure is obtained, missing mid \& high-frequency speech regions are filled in, and in the packet drop region, speech is naturally restored using contextual information.

\section{Conclusion}
We explored the speech restoration capability of voice conversion models in noisy conditions and demonstrated that NS speech output quality is significantly improved when combined with a voice conversion restoration module, while only incurring acceptable degradations in speech content intelligibility. This innovative approach addresses the limitations of traditional NS-only methods, offering a robust solution in the presence of multiple simultaneous distortions. To alleviate the degradation observed in the intelligibility metrics, we explored VC system content encoder adaptation to de-noised speech inputs and showed improved performance compared to the baseline VC version. We plan to investigate further noise/speech distortion adaptation methods, such as jointly training the noise suppression model, HuBERT, and ASR system, to further improve intelligibility while maintaining the high-quality speech observed with this 2-stage speech enhancement and restoration system.

\vfill\pagebreak
\bibliographystyle{IEEEtran}
\bibliography{main}

\begin{thebibliography}{10}
\providecommand{\url}[1]{#1}
\csname url@samestyle\endcsname
\providecommand{\newblock}{\relax}
\providecommand{\bibinfo}[2]{#2}
\providecommand{\BIBentrySTDinterwordspacing}{\spaceskip=0pt\relax}
\providecommand{\BIBentryALTinterwordstretchfactor}{4}
\providecommand{\BIBentryALTinterwordspacing}{\spaceskip=\fontdimen2\font plus
\BIBentryALTinterwordstretchfactor\fontdimen3\font minus \fontdimen4\font\relax}
\providecommand{\BIBforeignlanguage}[2]{{%
\expandafter\ifx\csname l@#1\endcsname\relax
\typeout{** WARNING: IEEEtran.bst: No hyphenation pattern has been}%
\typeout{** loaded for the language `#1'. Using the pattern for}%
\typeout{** the default language instead.}%
\else
\language=\csname l@#1\endcsname
\fi
#2}}
\providecommand{\BIBdecl}{\relax}
\BIBdecl

\bibitem{wang2018supervised}
D.~Wang and J.~Chen, ``Supervised speech separation based on deep learning: An overview,'' \emph{IEEE/ACM transactions on audio, speech, and language processing}, vol.~26, no.~10, pp. 1702--1726, 2018.

\bibitem{liu2022voicefixer}
H.~Liu, X.~Liu, Q.~Kong, Q.~Tian, Y.~Zhao, D.~Wang, C.~Huang, and Y.~Wang, ``Voicefixer: A unified framework for high-fidelity speech restoration,'' \emph{arXiv preprint arXiv:2204.05841}, 2022.

\bibitem{zhang2021restoring}
J.~Zhang, S.~Jayasuriya, and V.~Berisha, ``Restoring degraded speech via a modified diffusion model,'' \emph{arXiv preprint arXiv:2104.11347}, 2021.

\bibitem{goodfellow2020generative}
I.~Goodfellow, J.~Pouget-Abadie, M.~Mirza, B.~Xu, D.~Warde-Farley, S.~Ozair, A.~Courville, and Y.~Bengio, ``Generative adversarial networks,'' \emph{Communications of the ACM}, vol.~63, no.~11, pp. 139--144, 2020.

\bibitem{song2020score}
Y.~Song, J.~Sohl-Dickstein, D.~P. Kingma, A.~Kumar, S.~Ermon, and B.~Poole, ``Score-based generative modeling through stochastic differential equations,'' \emph{arXiv preprint arXiv:2011.13456}, 2020.

\bibitem{autovc}
K.~Qian, Y.~Zhang, S.~Chang, X.~Yang, and M.~Hasegawa-Johnson, ``Autovc: Zero-shot voice style transfer with only autoencoder loss,'' in \emph{International Conference on Machine Learning}.\hskip 1em plus 0.5em minus 0.4em\relax PMLR, 2019, pp. 5210--5219.

\bibitem{freevc}
J.~Li, W.~Tu, and L.~Xiao, ``Freevc: Towards high-quality text-free one-shot voice conversion,'' 2023.

\bibitem{kaneko2020cyclegan}
T.~Kaneko, H.~Kameoka, K.~Tanaka, and N.~Hojo, ``Cyclegan-vc3: Examining and improving cyclegan-vcs for mel-spectrogram conversion,'' \emph{arXiv preprint arXiv:2010.11672}, 2020.

\bibitem{diffvc}
V.~Popov, I.~Vovk, V.~Gogoryan, T.~Sadekova, M.~S. Kudinov, and J.~Wei, ``Diffusion-based voice conversion with fast maximum likelihood sampling scheme,'' in \emph{ICLR}, 2022.

\bibitem{borsos2023audiolm}
Z.~Borsos, R.~Marinier, D.~Vincent, E.~Kharitonov, O.~Pietquin, M.~Sharifi, D.~Roblek, O.~Teboul, D.~Grangier, M.~Tagliasacchi \emph{et~al.}, ``Audiolm: a language modeling approach to audio generation,'' \emph{IEEE/ACM transactions on audio, speech, and language processing}, vol.~31, pp. 2523--2533, 2023.

\bibitem{serra2022universal}
J.~Serr{\`a}, S.~Pascual, J.~Pons, R.~O. Araz, and D.~Scaini, ``Universal speech enhancement with score-based diffusion,'' \emph{arXiv preprint arXiv:2206.03065}, 2022.

\bibitem{gradtts}
V.~Popov, I.~Vovk, V.~Gogoryan, T.~Sadekova, and M.~Kudinov, ``Grad-tts: A diffusion probabilistic model for text-to-speech,'' in \emph{International Conference on Machine Learning}.\hskip 1em plus 0.5em minus 0.4em\relax PMLR, 2021, pp. 8599--8608.

\bibitem{hubert}
W.~Hsu, B.~Bolte, Y.~H. Tsai, K.~Lakhotia, R.~Salakhutdinov, and A.~Mohamed, ``Hubert: Self-supervised speech representation learning by masked prediction of hidden units,'' \emph{{IEEE} {ACM} Trans. Audio Speech Lang. Process.}, vol.~29, pp. 3451--3460, 2021.

\bibitem{devlin2018bert}
J.~Devlin, ``Bert: Pre-training of deep bidirectional transformers for language understanding,'' \emph{arXiv preprint arXiv:1810.04805}, 2018.

\bibitem{hifigan}
J.~Kong, J.~Kim, and J.~Bae, ``Hifi-gan: Generative adversarial networks for efficient and high fidelity speech synthesis,'' \emph{NeurIPS}, 2020.

\bibitem{guidance}
J.~Ho and T.~Salimans, ``Classifier-free diffusion guidance,'' \emph{CoRR}, vol. abs/2207.12598, 2022.

\bibitem{nisqa}
G.~Mittag, B.~Naderi, A.~Chehadi, and S.~M{\"{o}}ller, ``{NISQA:} {A} deep cnn-self-attention model for multidimensional speech quality prediction with crowdsourced datasets,'' in \emph{Interspeech}, 2021.

\bibitem{gordeeva2021meaning}
L.~Gordeeva, V.~Ershov, O.~Gulyaev, and I.~Kuralenok, ``Meaning error rate: Asr domain-specific metric framework,'' in \emph{Proceedings of the 27th ACM SIGKDD Conference on Knowledge Discovery \& Data Mining}, 2021, pp. 458--466.

\bibitem{taal2010short}
C.~H. Taal, R.~C. Hendriks, R.~Heusdens, and J.~Jensen, ``A short-time objective intelligibility measure for time-frequency weighted noisy speech,'' in \emph{2010 IEEE international conference on acoustics, speech and signal processing}.\hskip 1em plus 0.5em minus 0.4em\relax IEEE, 2010, pp. 4214--4217.

\bibitem{ecapa}
B.~Desplanques, J.~Thienpondt, and K.~Demuynck, ``{ECAPA-TDNN:} emphasized channel attention, propagation and aggregation in {TDNN} based speaker verification,'' in \emph{Interspeech}, 2020.

\bibitem{whisper}
A.~Radford, J.~W. Kim, T.~Xu, G.~Brockman, C.~McLeavey, and I.~Sutskever, ``Robust speech recognition via large-scale weak supervision,'' in \emph{International conference on machine learning}.\hskip 1em plus 0.5em minus 0.4em\relax PMLR, 2023, pp. 28\,492--28\,518.

\bibitem{libritts}
H.~Zen, V.~Dang, R.~Clark, Y.~Zhang, R.~J. Weiss, Y.~Jia, Z.~Chen, and Y.~Wu, ``Libritts: A corpus derived from librispeech for text-to-speech,'' in \emph{Interspeech}, 2019.

\bibitem{vctk}
C.~M.~K. Yamagishi, Junichi;~Veaux, ``Cstr vctk corpus: English multi-speaker corpus for cstr voice cloning toolkit (version 0.92), [sound]. university of edinburgh. the centre for speech technology research (cstr). https://doi.org/10.7488/ds/2645.''

\bibitem{knapp1976generalized}
C.~Knapp and G.~Carter, ``The generalized correlation method for estimation of time delay,'' \emph{IEEE transactions on acoustics, speech, and signal processing}, vol.~24, no.~4, pp. 320--327, 1976.

\bibitem{Dong2020}
\BIBentryALTinterwordspacing
X.~Dong and D.~S. Williamson, ``Towards real-world objective speech quality and intelligibility assessment using speech-enhancement residuals and convolutional long short-term memory networks,'' \emph{The Journal of the Acoustical Society of America}, vol. 148, no.~5, pp. 3348--3359, 2020. [Online]. Available: \url{https://pubs.aip.org/asa/jasa/article/148/5/3348/631860/Towards-real-world-objective-speech-quality-and}
\BIBentrySTDinterwordspacing

\bibitem{Taal2016}
\BIBentryALTinterwordspacing
C.~H. Taal, R.~C. Hendriks, R.~Heusdens, and J.~Jensen, ``An evaluation of objective measures for intelligibility prediction of time-frequency weighted noisy speech,'' \emph{PLOS ONE}, vol.~11, no.~3, p. e0150415, 2016. [Online]. Available: \url{https://journals.plos.org/plosone/article?id=10.1371/journal.pone.0150415}
\BIBentrySTDinterwordspacing

\end{thebibliography}

\end{document}